\documentclass[aps,preprint]{revtex4}
\usepackage{amsfonts}
\usepackage{amsmath}
\usepackage{amssymb}
\usepackage{graphicx}

\input{tcilatex}
\begin{document}

\preprint{}
\title[ ]{ Colliding Wave Solutions in a Symmetric Non-metric Theory.}
\author{Ozay Gurtug}
\email{ozay.gurtug@emu.edu.tr}
\affiliation{Department of Physics, Eastern Mediterranean University, G. Magusa, North
Cyprus, Mersin 10 - Turkey.}
\author{Mustafa Halilsoy}
\email{mustafa.halilsoy@emu.edu.tr}
\affiliation{Department of Physics, Eastern Mediterranean University, G. Magusa, North
Cyprus, Mersin 10 - Turkey.}
\pacs{PACS number}

\begin{abstract}
A method is given to generate the non-linear interaction (collision) of
linearly polarized gravity coupled torsion waves in a non-metric theory.
Explicit examples are given in which strong mutual focussing of
gravitational waves containing impulsive and shock components coupled with
torsion waves does not result in a curvature singularity. However, the
collision of purely torsion waves displays a curvature singularity in the
region of interaction.
\end{abstract}

\maketitle

\section{Introduction}

Horizon forming colliding plane wave (CPW) solutions in Einstein's general
relativity including Maxwell, scalar, dilaton, axion, Yang-Mills fields and
their various combinations have all been found so far\cite{JG},\cite{GHS},%
\cite{HH},\cite{HS},\cite{GH}. This amounts to finding solution that instead
of a curvature singularity analytically extendible Cauchy horizon forms in
the interaction region of colliding waves.

Being motivated by all these examples we wish to address in this paper to
the problem of whether similar type of horizon forming solutions can be
found in a non-metric theory of gravity. We achieve this goal indirectly,
namely by embedding particular non-metric theory into a metric one. For this
purpose we start with the Eisenhart's theory of unified fields in
Einstein-Cartan theory which upon reduction leads to the Einstein-scalar
(ES) theory. In a series of articles, Eisenhart attempted to unify
electromagnetism and gravity within the context of Einstein - Cartan theory 
\cite{E1},\cite{E2}. The choice of asymmetrical connection and its vanishing
Ricci tensor ( both symmetrical and asymmetrical parts) leads to conditions
that covers a variety of Einsteinian energy-momenta.

In this paper, we restrict ourselves entirely, through the choice of a
connection to a massless scalar field which is a particular form of the
Brans - Dicke - Jordan theory. From physical stand point the interesting
aspect in this approach \ is that the possible detection of torsion waves
amounts to detection of the scalar waves coupled with gravity.

In the second stage of our work, we convert solutions of colliding waves
obtained in one realm(i.e. ES theory) into solutions pertaining to the other
realm (i.e. non-metric theory). Hence, we choose some horizon forming
solutions of colliding ES plane waves and transform them into the non-metric
theory; thus by the dual interpretation we obtain solutions for colliding
gravitational waves containing impulsive and shock waves coupled with
torsion waves that lead to horizon forming metrics instead of singularities.
The significance of the obtained solution is not restricted to colliding
wave interpretation only, because the resulting metric, via the coordinate
transformations can also be interpreted to represent a non-singular
"distorted" Schwarzchild black-hole interior with scalar hair. We call it
distorted because the inclusion of the scalar field breaks the spherical
symmetry.

We also consider the collision of \ purely torsion waves in a non-metric
theory. However, the obtained solution displays curvature singularity as the
focussing hypersurface is approached. The problem of finding regular
solution in the context of purely scalar waves or purely torsion waves is
still open. In the literature, there are some singular CPW solutions in the
non-metric theory with torsion \cite{GA} and in the scalar tensor theories
without torsion \cite{BR}.

Our paper is organized as follows. In section II, we give the formalism
together with the main result which was obtained in Ref.\cite{GM} about the
equivalence of Einstein- scalar and Brans-Dicke-Jordan theories. In section
III, Einstein-scalar theory is considered. Within this context, we have
obtained massless scalar field extension of the horizon forming Yurtsever
(or independently Ferrari-Ibanez) solution. The regular character of the
solution is emphasized by calculating the Weyl and the Ricci scalars in
Appendix. Section IV, is devoted for the dual interpretation of the solution
obtained in section III. This is achieved by expressing the torsion wave and
non-metric tensor components. In section V, we consider the collision of
purely torsion waves in a non-metric theory. The paper is concluded with a
conclusion in section VI.

\section{The Formalism}

Long time ago, it was shown that \cite{GM}, Eisenhart's generalized
asymmetrical connection whose empty - space equations reproduce Einstein -
Maxwell field equations can be extended to cover the massless ES theory. The
main concern in such studies is to introduce torsion to the background
Riemann geometry by relaxing the metricity condition ,

\begin{equation}
\nabla _{\mu }g_{\alpha \beta }=0.
\end{equation}

The geometry in which the torsion is introduced is known as Einstein -
Cartan geometry defined by\cite{SCH},

\begin{equation}
\nabla _{\mu }g_{\alpha \beta }=-Q_{\mu \alpha \beta },
\end{equation}

where $\nabla _{\mu }$ represents the covariant derivative with respect to
the asymmetrical connection, and $Q_{\mu \alpha \beta }$ is a tensor which
measures the non-metricity with the property that $Q_{\mu \alpha \beta
}=Q_{\mu \beta \alpha }.$ The general asymmetrical connection which is
derived from Eq.(2) is,

\begin{equation}
\Gamma _{\mu \beta }^{\varkappa }=\QATOPD\{ \} {\varkappa }{\mu \beta
}+g^{\alpha \varkappa }\left( S_{\mu \beta \alpha }-S_{\mu \alpha \beta
}-S_{\beta \alpha \mu }\right) +\frac{1}{2}\left( Q_{\mu \beta }\text{ }%
^{\varkappa }+Q_{\beta \mu }\text{ }^{\varkappa }-Q_{\mu \beta }^{\varkappa
}\right)
\end{equation}

where $\QATOPD\{ \} {\varkappa }{\mu \beta }$ stands for the Christoffel
symbol. The torsion tensor $S_{\mu \beta }$ $^{\rho }$ can be found by
antisymmetrizing the connection given in Eq.(3) and is given by,

\begin{equation}
S_{\mu \beta }\text{ }^{\rho }=\frac{1}{2}\left( \Gamma _{\mu \beta }^{\rho
}-\Gamma _{\beta \mu }^{\rho }\right) =\Gamma _{\left[ \mu \beta \right]
}^{\rho },
\end{equation}

while the symmetrical component reads

\begin{equation}
\Gamma _{\left( \mu \beta \right) }^{\rho }=\frac{1}{2}\left( \Gamma _{\mu
\beta }^{\rho }+\Gamma _{\beta \mu }^{\rho }\right) .
\end{equation}

The contortion tensor $T_{\mu \beta }$ $^{\varkappa }$ can be constructed
from Eq.(3) in terms of torsion and non-metric tensors as,

\begin{equation}
T_{\mu \beta }\text{ }^{\varkappa }=\Gamma _{\mu \beta }^{\varkappa
}-\QATOPD\{ \} {\varkappa }{\mu \beta }=S_{\mu \beta \varkappa }-S_{\mu
\varkappa \beta }-S_{\beta \varkappa \mu }+\frac{1}{2}\left( Q_{\mu \beta
\varkappa }\text{ }+Q_{\beta \mu \varkappa }\text{ }-Q_{\varkappa \mu \beta
}\right)
\end{equation}

from which the following relations easily follow,

\begin{equation}
S_{\mu \beta \varkappa }=\frac{1}{2}\left( T_{\mu \beta \varkappa }-T_{\beta
\mu \varkappa }\right) ,
\end{equation}

and

\begin{equation}
Q_{\mu \beta \varkappa }=T_{\mu \beta \varkappa }+T_{\mu \varkappa \beta }.
\end{equation}

The Riemann and Ricci tensors of the generalized connection $\Gamma _{\mu
\beta }^{\varkappa }$ can be obtained by using the standard definitions
which are found as,

\begin{equation}
R_{\nu \mu \lambda }\text{ }^{\alpha }=K_{\nu \mu \lambda }\text{ }^{\alpha
}+T_{\mu \lambda }\text{ }_{;\nu }^{\alpha }-T_{\nu \lambda }\text{ }_{;\mu
}^{\alpha }+T_{\nu \rho }\text{ }^{\alpha }T_{\mu \lambda }\text{ }^{\rho
}-T_{\mu \rho }\text{ }^{\alpha }T_{\nu \lambda }\text{ }^{\rho }
\end{equation}

\begin{equation}
R_{\mu \lambda }=R_{\alpha \mu \lambda }\text{ }^{\alpha }=K_{\mu \nu
}+T_{\alpha \lambda }\text{ }_{;\nu }^{\alpha }-T_{\alpha \lambda }\text{ }%
_{;\mu }^{\alpha }+T_{\alpha \rho }\text{ }^{\alpha }T_{\mu \lambda }\text{ }%
^{\rho }-T_{\alpha \rho }\text{ }^{\alpha }T_{\nu \lambda }\text{ }^{\rho }
\end{equation}

where $K_{\nu \mu \lambda }$ $^{\alpha },K_{\mu \nu }$ are the Riemann and
Ricci tensors respectively and semicolon "$;$" is the covariant derivative
with respect to the Riemanian connection. It should be noted that the
obtained Ricci tensor $R_{\mu \nu }$ has an antisymmetric character such
that $R_{\mu \nu }\neq R_{\nu \mu }.$

In general, we assume that the contortion tensor $T_{\mu \alpha }$ $^{\beta
} $ is expressed in terms of the vectors $k_{\mu },l_{\mu \text{ \ }}$and $%
t_{\mu }$ which will be defined by

\begin{equation}
T_{\mu \alpha }\text{ }^{\beta }=g_{\mu \alpha }k^{\beta }+l_{\mu }\delta
_{\alpha }^{\beta }+t_{\alpha }\delta _{\mu }^{\beta }.
\end{equation}

Substituting this into Eq. (10) will yield,

\begin{equation}
R_{\mu \nu }=K_{\mu \lambda }+g_{\mu \lambda }\left( k_{;\alpha }^{\alpha
}+k^{2}+3t\cdot k\right) +l_{\mu ;\lambda }-l_{\lambda ;\mu }-k_{\lambda
;\mu }-3t_{\lambda ;\mu }+3t_{\lambda }t_{\mu }-k_{\mu }k_{\lambda },
\end{equation}

whose symmetric and antisymmetric components become, respectively

\begin{equation}
R_{\left( \mu \nu \right) }=K_{\mu \lambda }+g_{\mu \lambda }\left(
k_{;\alpha }^{\alpha }+k^{2}+3t\cdot k\right) -\frac{1}{2}\left( k_{\lambda
;\mu }+k_{\mu ;\lambda }\right) -\frac{3}{2}\left( t_{\lambda ;\mu }+t_{\mu
;\lambda }\right) +3t_{\lambda }t_{\mu }-k_{\mu }k_{\lambda },
\end{equation}

\begin{equation}
R_{\left[ \mu \lambda \right] }=l_{\mu ;\lambda }-l_{\lambda ;\mu }-\frac{1}{%
2}\left( k_{\lambda ;\mu }-k_{\mu ;\lambda }\right) -\frac{3}{2}\left(
t_{\lambda ;\mu }-t_{\mu ;\lambda }\right) .
\end{equation}

Let $\phi $ be a scalar field and define the vector $l_{\mu }$ as

\begin{equation}
l_{\mu }=\phi _{,\mu }-\frac{1}{2}\left( 3t_{\mu }+k_{\mu }\right) ,
\end{equation}

which satisfies $R_{\left[ \mu \lambda \right] }=0.$ Automatically the other
vectors $t_{\mu }$ and $k_{\mu }$ are obtained from the symmetric component,
with the condition $R_{\left( \mu \nu \right) }=0.$ A variety of Einstein's
equations with sources can be obtained if one makes the choice as

\begin{equation}
\left( 3t_{\lambda }+k_{\lambda }\right) _{;\mu }=3t_{\mu }t_{\lambda
}-k_{\mu }k_{\lambda }+\omega T_{\mu \lambda }+fg_{\mu \nu }.
\end{equation}

Here $\omega $ and $f$ are functions to be found, $T_{\mu \lambda }$ is the
symmetric energy momentum tensor to be specified. If the above relation is
substituted in $R_{\left( \mu \nu \right) }=0,$ with the choice of the
function $f$ as

\begin{eqnarray}
f &=&3t\cdot k+k^{2}+k_{;\alpha }^{\alpha }-\frac{\omega }{2}T, \\
T &\equiv &T_{\alpha }^{\alpha },  \notag
\end{eqnarray}

the following result will be obtained,

\begin{equation}
K_{\mu \nu }=\omega \left( T_{\mu \nu }-\frac{1}{2}Tg_{\mu \lambda }\right) .
\end{equation}

This is exactly the Einstein's equations with sources which can be written as

\begin{equation}
G_{\mu \nu }\equiv K_{\mu \nu }-\frac{1}{2}Kg_{\mu \lambda }=\omega T_{\mu
\nu },
\end{equation}

while $\omega $ becomes the coupling constant.

As a particular example let us consider the coupling of a massless scalar
field $\phi $. The vectors are taken as follow,

\begin{equation}
l_{\mu }=-t_{\mu }=\frac{1}{3}k_{\mu }=\sqrt{\frac{\kappa }{6}}\phi _{,\mu },
\end{equation}

where $\kappa $ is a constant. The contortion tensor is defined by ,

\begin{equation}
T_{\mu \nu }\text{ }^{\beta }=\sqrt{\frac{\kappa }{6}}\left( 3g^{\beta
\gamma }\phi _{,\gamma }g_{\mu \nu }+\delta _{\nu }^{\beta }\phi _{,\mu
}-\delta _{\mu }^{\beta }\phi _{,\nu }\right)
\end{equation}

with the trivial constraint condition $T_{\alpha \lambda }$ $^{\alpha }=0,$
and the non-trivial one $T_{\mu \nu }$ $_{;\alpha }^{\alpha }=0,$ which is
equivalent to the ES field equation $g^{\alpha \beta }\phi _{;\alpha \beta
}=0.$

Hence the Ricci tensor becomes

\begin{equation}
R_{\mu \nu }\equiv K_{\mu \nu }-\kappa \phi _{,\mu }\phi _{,\nu }=0,
\end{equation}

which yields

\begin{equation}
K_{\mu \nu }=\kappa \phi _{,\mu }\phi _{,\nu },
\end{equation}

which is the Einstein - massless scalar field equations.

According to this choice the torsion and non-metric tensors are defined as,

\begin{equation}
S_{\mu \beta \varkappa }=\sqrt{\frac{\kappa }{6}}\left( g_{\beta \varkappa
}\phi _{,\mu }-g_{\mu \varkappa }\phi _{,\beta }\right) ,
\end{equation}

\begin{equation}
Q_{\mu \beta \varkappa }=2\sqrt{\frac{\kappa }{6}}\left( g_{\mu \beta }\phi
_{,\varkappa }+g_{\beta \varkappa }\phi _{,\mu }+g_{\mu \varkappa }\phi
_{,\beta }\right) .
\end{equation}

\section{ Solution For Colliding ES Waves.}

The adopted space-time line element in general for linearly polarized case
is given in Szekeres form by,

\begin{equation}
ds^{2}=2e^{-M}dudv-e^{-U}\left( e^{V}dx^{2}+e^{-V}dy^{2}\right) .
\end{equation}

The metric functions $M,U$ and $V$ are functions of the null coordinates $u$
and $v$ only. In Ref.\cite{GHS}, the field equations are derived for the
problem of colliding Einstein-Maxwell-scalar waves for non-linearly
polarized waves and the \textit{M-shift} method is explained how to extend
the vacuum (Einstein) or electrovacuum (Einstein-Maxwell) solutions to a
vacuum (electrovacuum)-scalar solutions.

In this paper, we shall use the same field equations which is shown in the
previous section that they are equivalent to the field equations of
Brans-Dicke-Jordan theory for Einstein-scalar case to obtain a class of
regular solutions which represents colliding gravitational waves in the
symmetric, non-metric space-times with torsion.

As a requirement of the \textit{M-shift} method, the scalar field $\phi $ is
coupled to gravitational wave through shifting the metric function $M$ in
Eq.(13), in Ref.\cite{GHS} (see Ref.\cite{GHS} for details) in accordance
with,

\begin{equation}
M\rightarrow \widetilde{M}=M+\Gamma
\end{equation}

where the function $\Gamma $ derives from the the presence of the scalar
field $\phi ,$ through the conditions

\begin{equation}
U_{u}\Gamma _{u}=2\phi _{u}^{2}\text{ \ \ \ \ \ \ \ \ \ \ and \ \ \ \ \ \ \
\ }U_{v}\Gamma _{v}=2\phi _{v}^{2}.
\end{equation}

We note that throughout the paper a subscript notation implies partial
derivative. The integrability condition induces the massless scalar field
equation as a constraint condition,

\begin{equation}
2\phi _{uv}-U_{u}\phi _{v}-U_{v}\phi _{u}=0.
\end{equation}

The most general solution to this equation is obtained if the prolate type
of coordinates $\left( \tau ,\sigma \right) $ is used instead of the null
coordinates $\left( u,v\right) .$ The relation between these coordinates are
defined by,

\begin{equation}
\tau =\sin \left( au_{+}+bv_{+}\right) ,\text{ \ \ \ \ \ \ \ \ \ \ \ \ \ \ \ 
}\sigma =\sin \left( au_{+}-bv_{+}\right) ,
\end{equation}

where $a,b$ are constants and $u_{+}=u\theta \left( u\right) ,$ $%
v_{+}=v\theta \left( v\right) $ with $\theta \left( u\right) $ and $\theta
\left( v\right) $ are unit step functions. In terms of prolate coordinates
the massless scalar field equation (29) and conditions (28) becomes,

\begin{equation}
\left( \Delta \phi _{\tau }\right) _{\tau }-\left( \delta \phi _{\sigma
}\right) _{\sigma }=0,
\end{equation}

\begin{eqnarray}
\left( \tau ^{2}-\sigma ^{2}\right) \Gamma _{\tau } &=&2\Delta \delta \left(
\tau \phi _{\tau }^{2}+\frac{\tau \delta }{\Delta }\phi _{\sigma
}^{2}-2\sigma \phi _{\tau }\phi _{\sigma }\right) , \\
\left( \sigma ^{2}-\tau ^{2}\right) \Gamma _{\sigma } &=&2\Delta \delta
\left( \sigma \phi _{\sigma }^{2}+\frac{\sigma \Delta }{\delta }\phi _{\tau
}^{2}-2\tau \phi _{\tau }\phi _{\sigma }\right) ,  \notag
\end{eqnarray}

where $\Delta =1-\tau ^{2}$ and $\delta =1-\sigma ^{2}.$ The exact solution
to Eq.(31) is already available in \cite{JG};

\begin{equation}
\phi \left( \tau ,\sigma \right) =\sum_{n}\left\{ a_{n}P_{n}(\tau
)P_{n}(\sigma )+b_{n}Q_{n}(\tau )Q_{n}(\sigma )+c_{n}P_{n}(\tau
)Q_{n}(\sigma )+d_{n}P_{n}(\sigma )Q_{n}(\tau )\right\} ,
\end{equation}

where $P$ and $Q$ are the Legendre functions of the first and second kind
respectively, and $a_{n},$ $b_{n},$ $c_{n}$ and $d_{n}$ are arbitrary
constants. The choice of scalar field $\phi \left( \tau ,\sigma \right) $ is
extremly important as far as the regular and physically acceptable solutions
are concerned. The regular solutions will be obtained if $%
b_{n}=c_{n}=d_{n}=0 $ and $a_{n}\neq 0.$ We choose the scalar field as,

\begin{equation}
\phi \left( \tau ,\sigma \right) =\alpha \tau \sigma +\frac{1}{4}\beta
\left( 3\tau ^{2}-1\right) \left( 3\sigma ^{2}-1\right)
\end{equation}

where $\alpha $ and $\beta $ are arbitrary constants. Integration of Eq.(32)
yields

\begin{equation}
\Gamma =\alpha ^{2}\left( \tau ^{2}+\sigma ^{2}\Delta \right) +\frac{9}{8}%
\beta ^{2}\left[ \Delta \left( \tau ^{2}+\sigma ^{2}\Delta \right) +\tau ^{2}%
\right] -6\alpha \beta \tau \sigma \Delta \delta .
\end{equation}

For the problem at hand, we couple the scalar field to the Cauchy - horizon
forming pure gravitational wave solution obtained long ago by Yurtsever\cite%
{YU} \ ( or independently Ferrari-Ibanez\cite{FI}) . This particular
solution is known to be isometric to the part of interior region of the
Schwarzchild black hole.

The resulting metric that describes the collision of plane impulsive waves
accompanied by shock gravitational waves coupled with massless scalar field
is given by

\begin{equation}
ds^{2}=2e^{-\widetilde{M}}dudv-e^{-U}\left( e^{V}dx^{2}+e^{-V}dy^{2}\right) ,
\end{equation}

where the metric functions are,

\begin{eqnarray}
e^{-\widetilde{M}} &=&\left( 1+\tau \right) ^{2}e^{-\Gamma }, \\
e^{-V} &=&\sqrt{\frac{\delta }{\Delta }}\left( 1+\tau \right) ^{2},  \notag
\\
e^{-U} &=&\sqrt{\Delta \delta }.  \notag
\end{eqnarray}

We have shown with this example that, it is possible to construct a class of
exact colliding parallel polarized plane wave solutions in the
Einstein-scalar theory. Among the others it is shown that when a particular
type of scalar fields couples as an initial data to an incoming parallel
polarized gravitational wave results a non-singular Cauchy-horizon in the
interaction region. This regularity is clearly evident by analysing the Weyl
and Ricci scalars which are given in Appendix.

Another physical interpretation of the Eq.36 is possible if the following
coordinate transformation is used. Let $\psi =au_{+}+bv_{+}$ and $\lambda
=au_{+}-bv_{+}$, together with

\begin{equation}
r=1+\sin \psi ,\text{ \ \ \ \ }\theta =\frac{\pi }{2}-\lambda ,\text{ \ \ }\
\ t=\sqrt{2}x,\text{ \ \ \ }\varphi =\sqrt{2}y,
\end{equation}

transforms the line element ( Eq.36) into,

\begin{equation}
ds^{2}=(1-\frac{2}{r})dt^{2}-e^{-\Gamma }(1-\frac{2}{r}%
)^{-1}dr^{2}-r^{2}[e^{-\Gamma }d\theta ^{2}+\sin ^{2}\theta d\varphi ^{2}],
\end{equation}

in which the range of the coordinate $0\leq \psi \leq \frac{\pi }{2}$ ,
confines the radial coordinate to $1\leq r\leq 2.$ In the absence of the
scalar field, the metric (Eq.39) corresponds to the Schwarzchild black -
hole interior with mass $m=1.$ \ With the scalar field, the metric is no
more spherically symmetric and could be interpreted to represent distorted
Schwarzchild black-hole interior with scalar hair.

\section{Colliding Waves in a non-metric Theory.}

In section II, an analogy has been established between the Einstein-scalar
and non-metric theories. As an outcome of this analogy, we consider the
non-linear interaction (collision) of gravitational waves with torsion. The
waves that participates in the collision are parallely polarized impulsive
gravitational waves accompanied with shock gravitational waves coupled with
torsion waves.

In general, the whole spacetime is divided into four continuous regions with
the appropriate boundary conditions. These four regions are depicted in
Fig.1. Region I ($u<0,v<0$), is the flat Minkowski region: Region II ( $%
u>0,v<0$) and Region III ( $u<0,v>0$) are the plane symmetric incoming
regions that contains the waves participating in the collision: Region IV ( $%
u>0,v>0$) is the interaction region.

The metric and metric functions that describes the collision of parallely
polarized impulsive gravitational waves accompanied with shock gravitational
waves coupled with torsion waves are given in Eq.(36) and (37) respectively.
The torsion waves (which is purely tensor) and the non-metric tensor
components are evaluated by using the equations (24) and (25) respectively
and given by

\begin{eqnarray}
S_{uvu} &=&\sqrt{\frac{\kappa }{6}}e^{-M}\phi _{u},\text{ \ \ \ \ }S_{uxx}=-%
\sqrt{\frac{\kappa }{6}}\frac{\Delta }{\left( 1+\tau \right) ^{2}}\phi _{u},%
\text{ \ \ \ \ }S_{uyy}=-\sqrt{\frac{\kappa }{6}}\delta \left( 1+\tau
\right) ^{2}\phi _{u}, \\
S_{uvv} &=&-\sqrt{\frac{\kappa }{6}}e^{-M}\phi _{v},\text{ \ \ }S_{vxx}=-%
\sqrt{\frac{\kappa }{6}}\frac{\Delta }{\left( 1+\tau \right) ^{2}}\phi _{v},%
\text{ \ \ \ \ }S_{vyy}=-\sqrt{\frac{\kappa }{6}}\delta \left( 1+\tau
\right) ^{2}\phi _{v},  \notag
\end{eqnarray}

and

\begin{eqnarray}
Q_{uuv} &=&4\sqrt{\frac{\kappa }{6}}e^{-M}\phi _{u},\text{ \ \ \ }Q_{uxx}=-%
\sqrt{\frac{2\kappa }{3}}\frac{\Delta }{\left( 1+\tau \right) ^{2}}\phi _{u},%
\text{\ \ \ }Q_{uyy}=-\sqrt{\frac{2\kappa }{3}}\delta \left( 1+\tau \right)
^{2}\phi _{u},\text{\ } \\
Q_{vvu} &=&4\sqrt{\frac{\kappa }{6}}e^{-M}\phi _{v},\text{ \ \ \ }Q_{vxx}=-%
\sqrt{\frac{2\kappa }{3}}\frac{\Delta }{\left( 1+\tau \right) ^{2}}\phi _{v},%
\text{ \ \ \ }Q_{vyy}=-\sqrt{\frac{2\kappa }{3}}\delta \left( 1+\tau \right)
^{2}\phi _{v},  \notag
\end{eqnarray}

where $\phi _{u}$ and $\phi _{v}$ are,

\begin{eqnarray*}
\phi _{u} &=&a\left\{ \alpha \sin 2au_{+}+\frac{3}{2}\beta \left[ -\frac{1}{2%
}\left( \sin 2\psi +\sin 2\lambda \right) +3\sin \lambda \sin \psi \sin
2au_{+}\right] \right\} \theta \left( u\right) , \\
\phi _{v} &=&b\left\{ -\alpha \sin 2bv_{+}+\frac{3}{2}\beta \left[ \frac{1}{2%
}\left( \sin 2\lambda -\sin 2\psi \right) -3\sin \lambda \sin \psi \sin
2bv_{+}\right] \right\} \theta \left( v\right) .
\end{eqnarray*}

\section{Colliding Purely Torsion Waves.}

In this section, we consider the collision of linearly polarized purely
torsion waves in a non-metric theory. The analogous problem of colliding
complex and real massless scalar waves in the Einstein theory was considered
long ago in the references \cite{WU} and \cite{HA} respectively. This is
accomplished by taking the metric function $V=0$ in Eq.(26) and the line
element becomes

\begin{equation}
ds^{2}=2e^{-M}dudv-e^{-U}\left( dx^{2}+dy^{2}\right) .
\end{equation}

This choice renders all the Weyl scalars to vanish in the incoming regions.
So that the non-vanishing Ricci scalars in these regions implies purely
torsion waves. Another consequence of this choice is the simplifications in
the field equations describing the collision of purely torsion waves which
are given by,

\begin{equation}
U_{uv}=U_{u}U_{v}-2\Phi _{11}^{\left( 0\right) }-6\Lambda ^{\left( 0\right) }
\end{equation}

\begin{equation}
2U_{uu}=U_{u}^{2}-2U_{u}M_{u}+4\phi _{u}^{2},
\end{equation}

\begin{equation}
2U_{vv}=U_{v}^{2}-2U_{v}M_{v}+4\phi _{v}^{2},
\end{equation}

\begin{equation}
2M_{uv}=-U_{u}U_{v}+4\phi _{u}\phi _{v,}
\end{equation}

and the massless-scalar field equation which becomes equivalent to the
source of the torsion waves in \ a non-metric theory is given by,

\begin{equation}
2\phi _{uv}=U_{u}\phi _{v}+U_{v}\phi _{u}.
\end{equation}

It has been found more convinient to use prolate type coordinates in
obtaining solutions to Eq.(47) . Using the following transformations

\begin{eqnarray}
\tau &=&u\sqrt{1-v^{2}}+v\sqrt{1-u^{2}}, \\
\sigma &=&u\sqrt{1-v^{2}}-v\sqrt{1-u^{2}},  \notag
\end{eqnarray}

\qquad\ the Eq(47) transforms into Eq(31). One of the solution to Eq.(31) is
the Szekeres solution that guarantees to satisfy the boundary conditions. In
prolate spheroidal coordinates this is given by,

\begin{equation}
\phi \left( u,v\right) =\frac{1}{2}\ln \left( \frac{1+\tau }{1-\tau }\right)
.
\end{equation}

The resulting solution is obtained as,

\begin{equation*}
e^{-U}=1-u^{2}-v^{2}
\end{equation*}

\begin{equation}
e^{-M}=\frac{\left( 1-u^{2}-v^{2}\right) ^{3/2}}{\sqrt{1-u^{2}}\sqrt{1-v^{2}}%
\left( uv+\sqrt{1-u^{2}}\sqrt{1-v^{2}}\right) ^{2}}.
\end{equation}

The non zero torsion waves and non-metric tensor components are,

-%
\begin{eqnarray}
S_{uvu} &=&\sqrt{\frac{\kappa }{6}}e^{-M}\phi _{u},\text{ \ \ \ \ \ }%
S_{uxx}=-\sqrt{\frac{\kappa }{6}}e^{-U}\phi _{u},\text{ \ \ \ \ }S_{uyy}=-%
\sqrt{\frac{\kappa }{6}}e^{-U}\phi _{u}, \\
S_{uvv} &=&-\sqrt{\frac{\kappa }{6}}e^{-M}\phi _{v},\text{ \ \ \ }S_{vxx}=-%
\sqrt{\frac{\kappa }{6}}e^{-U}\phi _{v},\text{ \ \ \ }S_{vyy}=-\sqrt{\frac{%
\kappa }{6}}e^{-U}\phi _{v},  \notag
\end{eqnarray}

and

\begin{eqnarray}
Q_{uuv} &=&4\sqrt{\frac{\kappa }{6}}e^{-M}\phi _{u},\text{ \ \ \ \ \ }%
Q_{uxx}=-\sqrt{\frac{2\kappa }{3}}e^{-U}\phi _{u},\text{ \ \ \ }Q_{uyy}=-%
\sqrt{\frac{2\kappa }{3}}e^{-U}\phi _{u}, \\
Q_{vvu} &=&4\sqrt{\frac{\kappa }{6}}e^{-M}\phi _{v},\text{ \ \ \ \ \ }%
Q_{vxx}=-\sqrt{\frac{2\kappa }{3}}e^{-U}\phi _{v},\text{ \ \ \ }Q_{vyy}=-%
\sqrt{\frac{2\kappa }{3}}e^{-U}\phi _{v},  \notag
\end{eqnarray}

where

\begin{eqnarray}
\phi _{u} &=&\frac{\theta \left( u\right) }{\sqrt{1-u^{2}}\left( \sqrt{%
1-u^{2}}\sqrt{1-v^{2}}-uv\right) } \\
\phi _{v} &=&\frac{\theta \left( v\right) }{\sqrt{1-v^{2}}\left( \sqrt{%
1-u^{2}}\sqrt{1-v^{2}}-uv\right) }  \notag
\end{eqnarray}

We note that the null coordinates $u$ and $v$ are implied with a step
functions $u\rightarrow u\theta \left( u\right) $ and $v\rightarrow v\theta
\left( v\right) $ respectively. In contrast to the gravity coupled torsion
waves, this particular example exhibits curvature singularity as the
focussing hypersurface $u^{2}+v^{2}\rightarrow 1$ is approached. This is
indicated in the scale invariant Weyl \ scalar $\Psi _{2}^{\left( 0\right) }$
that arises as a result of non-linear interaction in region IV,

\begin{equation}
\Psi _{2}^{\left( 0\right) }=\frac{\left( \sqrt{1-u^{2}}\sqrt{1-v^{2}}%
+uv\right) ^{2}}{\sqrt{1-u^{2}}\sqrt{1-v^{2}}\left( 1-u^{2}-v^{2}\right) ^{2}%
}-\frac{uv}{\left( 1-u^{2}-v^{2}\right) ^{2}}.
\end{equation}

The non-zero Ricci scalars are,

\begin{eqnarray}
\Phi _{00}^{\left( 0\right) } &=&\frac{\theta \left( v\right) }{\left(
1-u^{2}\right) \left( \sqrt{1-u^{2}}\sqrt{1-v^{2}}-uv\right) ^{2}}, \\
\Phi _{22}^{\left( 0\right) } &=&\frac{\theta \left( u\right) }{\left(
1-v^{2}\right) \left( \sqrt{1-u^{2}}\sqrt{1-v^{2}}-uv\right) ^{2}},  \notag
\\
\Phi _{11}^{\left( 0\right) } &=&\frac{\theta \left( u\right) \theta \left(
v\right) }{\sqrt{1-u^{2}}\sqrt{1-v^{2}}\left( \sqrt{1-u^{2}}\sqrt{1-v^{2}}%
-uv\right) ^{2}},  \notag \\
\Lambda ^{\left( 0\right) } &=&-\frac{1}{3}\Phi _{11}^{\left( 0\right) }. 
\notag
\end{eqnarray}

\section{Conclusion.}

In this study, we have presented two types of colliding plane wave solutions
in the symmetric non-metric theory. This is accomplished by using an analogy
which was developed long ago between the metric and non-metric theories.
This analogy reveals the equivalence of the field equations in
Einstein-scalar and Brans-Dicke-Jordan theories.

One of the obtained solution describes the collision of impulsive
gravitational waves accompanied with shock gravitational waves coupled with
torsion waves. This particular solution has an interesting property that, in
the region of interaction,\ an analytically extendible Cauchy horizon forms
in place of a curvature singularity. On the other hand, the collision of
purely torsion waves results in a curvatu\i re singularity in the
interaction region.

\textbf{APPENDIX}

The non-zero Weyl and Ricci scalars for the collision of parallely polarized
impulsive gravitational waves accompanied with shock gravitational waves
coupled with massless-scalar wave are obtained as follows.

\bigskip

\begin{eqnarray}
\Psi _{2} &=&\frac{9abe^{\Gamma }}{8\left( 1+\sin \psi \right) ^{3}}\{\frac{%
\beta ^{2}\cos ^{2}\psi }{3}\left[ \cos ^{2}\psi +3\cos ^{2}\lambda \left(
\cos 2\psi -1\right) \right] \left( 1+\sin \psi \right) + \\
&&\frac{4}{9}\alpha \left( 1+\sin \psi \right) \left( \cos 2\psi -\cos
2\lambda \right) \left[ 2\beta \sin \psi \sin \lambda +\frac{\alpha }{3}%
\right] +\frac{8}{9}\}\theta \left( u\right) \theta \left( v\right) ,  \notag
\end{eqnarray}

\begin{eqnarray}
e^{-\Gamma }\Psi _{0} &=&\frac{b}{\left( 1+\sin au\right) ^{2}\cos au}\delta
\left( v\right) +\frac{2b^{2}\theta \left( v\right) }{\left( 1+\sin \psi
\right) ^{2}}\{\frac{9\beta ^{2}\cos ^{2}\psi }{16}[\cos ^{2}\psi \left(
3\cos ^{2}\lambda -1\right) + \\
&&\cos \psi \cos \lambda \sin \lambda \left( 3\sin \psi -2\right) +2\cos
^{2}\lambda \left( 2\sin \psi -1\right) ]+\frac{3}{2}\alpha \beta \{\cos
\psi \cos \lambda \lbrack 4+  \notag \\
&&\cos ^{2}\psi \left( 6\cos ^{2}\lambda -5\right) +\sin \psi \left( 6\cos
^{2}\lambda -4\right) -5\cos ^{2}\lambda ]+  \notag \\
&&2\sin \lambda \cos ^{2}\lambda \lbrack 3\cos ^{2}\psi \left( \sin \psi
-1\right) -\sin \psi +2]-2\sin \psi \sin \lambda \cos ^{2}\psi \}+  \notag \\
&&\frac{\alpha ^{2}}{2}[2\cos \psi \sin \lambda \cos \lambda \left( \sin
P-1\right) +\cos ^{2}\lambda \left( 2\sin \psi -1\right) +\cos ^{2}\psi
\left( 2\cos ^{2}\lambda -1\right) ]-  \notag \\
&&\frac{3}{2}\left( 1+\sin \psi \right) ^{-1}\},  \notag
\end{eqnarray}

\begin{eqnarray}
e^{-\Gamma }\Psi _{4} &=&\frac{a}{\left( 1+\sin bv\right) ^{2}\cos bv}\delta
\left( u\right) -\frac{2a^{2}\theta \left( u\right) }{\left( 1+\sin \psi
\right) ^{2}}\{\frac{9\beta ^{2}\cos ^{2}\psi }{16}[\cos ^{2}\psi \left(
1-3\cos ^{2}\lambda \right) + \\
&&\cos \psi \cos \lambda \sin \lambda \left( 3\sin \psi -2\right) +2\cos
^{2}\lambda \left( 1-2\sin \psi \right) ]+\frac{3}{2}\alpha \beta \{\cos
\psi \cos \lambda \lbrack 4+  \notag \\
&&\cos ^{2}\psi \left( 6\cos ^{2}\lambda -5\right) +\sin \psi \left( 6\cos
^{2}\lambda -4\right) -5\cos ^{2}\lambda ]+  \notag \\
&&2\sin \lambda \cos ^{2}\lambda \lbrack 3\cos ^{2}\psi \left( 1-\sin \psi
\right) +\sin \psi -2]+2\sin \psi \sin \lambda \cos ^{2}\psi \}+  \notag \\
&&\frac{\alpha ^{2}}{2}[2\cos \psi \sin \lambda \cos \lambda \left( \sin
\psi -1\right) +\cos ^{2}\lambda \left( 1-2\sin \psi \right) +\cos ^{2}\psi
\left( 1-2\cos ^{2}\lambda \right) ]+  \notag \\
&&\frac{3}{2}\left( 1+\sin \psi \right) ^{-1}\},  \notag
\end{eqnarray}

\bigskip

\begin{eqnarray}
e^{-\Gamma }\Phi _{22} &=&\frac{2a^{2}}{\left( 1+\sin \psi \right) ^{2}}\{%
\frac{9\beta ^{2}\cos ^{2}\psi \sin 2au}{16}[\sin 2au+\sin \psi \cos \lambda
] \\
&&+\frac{3}{2}\alpha \beta \{2\cos 2au[\cos ^{2}\psi \left( 2\cos
^{2}\lambda -1\right) -\cos ^{2}\lambda \sin ^{2}\psi ]  \notag \\
&&+3\cos \psi \cos \lambda \left( \frac{4}{3}-\cos ^{2}\lambda -\cos
^{2}\psi \right) \}+\frac{\alpha ^{2}}{2}\sin ^{2}2au\}  \notag
\end{eqnarray}

\bigskip

\begin{eqnarray}
e^{-\Gamma }\Phi _{00} &=&\frac{2b^{2}}{\left( 1+\sin \psi \right) ^{2}}\{%
\frac{9\beta ^{2}\cos ^{2}\psi \sin 2bv}{16}[\sin 2bv+\sin \psi \cos \lambda
] \\
&&+\frac{3}{2}\alpha \beta \{2\cos 2bv[\cos ^{2}\lambda \left( 2\cos
^{2}\psi -1\right) -\cos ^{2}\psi \sin ^{2}\lambda ]  \notag \\
&&+3\cos \psi \cos \lambda \left( \frac{4}{3}-\cos ^{2}\lambda -\cos
^{2}\psi \right) \}+\frac{\alpha ^{2}}{2}\sin ^{2}2au\}  \notag
\end{eqnarray}

\bigskip

\begin{equation}
\Phi _{02}=0,
\end{equation}

\bigskip

\begin{eqnarray}
\Phi _{11} &=&-3\Lambda =\frac{abe^{\Gamma }}{16\left( 1+\sin \psi \right)
^{2}}\{\beta ^{2}\{9\cos ^{2}\psi \lbrack 3\cos ^{2}\lambda \left( \cos
2\psi -1\right) +\cos ^{2}\psi ]\}- \\
&&4\alpha \left( \cos 2\lambda -\cos 2\psi \right) \left( 6\beta \sin \psi
\sin \lambda +\alpha \right) \}\theta \left( u\right) \theta \left( v\right)
.  \notag
\end{eqnarray}

\begin{center}
\bigskip FIGURE CAPTION
\end{center}

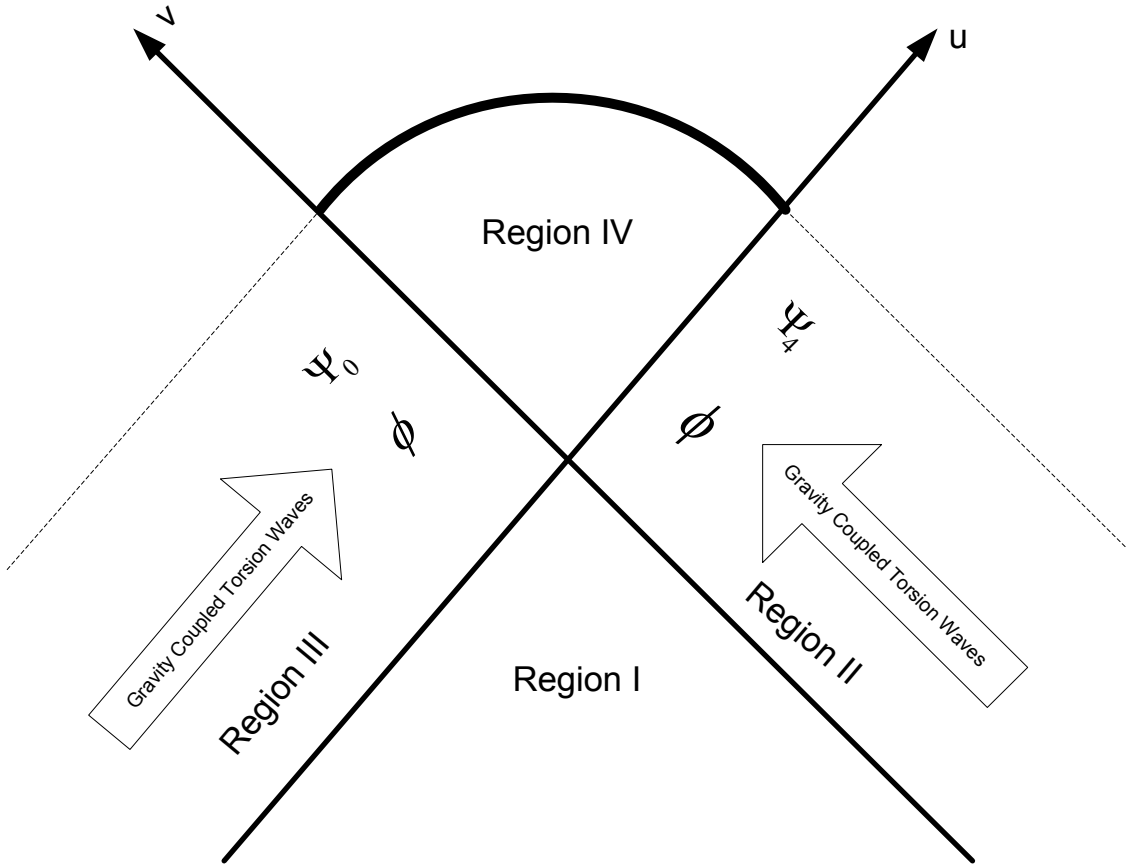
Figure 1: The space-time diagram describes the collision of gravitational
waves coupled with torsion waves.

\end{document}